\documentclass{ws-ijmpe}
\usepackage{url}
\usepackage[super,compress]{cite}

\usepackage{hyperref}
\hypersetup{colorlinks,urlcolor=black,citecolor=black,linkcolor=black,filecolor=black}
\usepackage{breakurl}
\usepackage{color}
\usepackage{outlines}
\begin{document}

\markboth{M.A. Famiano}{Nuclear Mass Measurements With Radioactive Ion Beams}

\catchline{}{}{}{}{}

\title{Nuclear Mass Measurements With Radioactive Ion Beams}

\author{Michael A. Famiano}

\address{Physics Department, Western Michigan University\\
	1903 W. Michigan Ave.\\
	Kalamazoo, MI 49008-5252 United States of America\\
	michael.famiano@wmich.edu}

\maketitle

\begin{history}
\received{X XXXXX XXXX}
\revised{XX XXXXX XXXX}
\end{history}

\begin{abstract}
	Nuclear masses are the most fundamental of all nuclear properties, yet
	they can provide a wealth of knowledge, including information on
	astrophysical sites, constraints on existing theory, and 
	fundamental symmetries.  In nearly all applications, it is necessary to
	measure nuclear masses with very high precision.  As mass measurements
	push to more short-lived and more massive nuclei, the practical constraints
	on mass measurement techniques become more exacting.  Various
	techniques used to measure nuclear masses, including their 
	advantages and disadvantages are described.  Descriptions of
	some of the 
	world facilities at which the nuclear mass measurements are performed 
	are given, and brief summaries of planned facilities are presented.  Future directions
	are mentioned, and conclusions are presented which provide a possible outlook
	and emphasis on upcoming plans for nuclear mass measurements at existing
	facilities, those under construction, and those being planned.
\end{abstract}


\ccode{PACS numbers:21.60.-n; 26.30.-k; 29.27.-a}

\tableofcontents
\section{Introduction}
Besides proof of particle stability, perhaps the most fundamental of nuclear properties is nuclear mass.  Knowledge of nuclear
masses can provide significant insight into other nuclear properties including, but not limited to, particle separation energies,
decay rates, shape, and shell model predictions.  Because of their intrinsic relationship to a host of other nuclear properties,
theoretical predictions and experimental determinations of nuclear masses have been a mainstay of nuclear research virtually since
its inception.  

Multiple techniques
for measuring nuclear mass have evolved over the years, each with characteristic advantages and disadvantages. In many cases,
the disadvantages of one particular technique are an advantage of another.  Applications of nuclear mass measurements to fundamental
nuclear physics and astrophysics will be discussed.  Mass measurement techniques will also be discussed along with the
facilities at which various methods are used.  Finally, future directions
in nuclear mass measurements will be presented.
\subsection{Applications of Nuclear Masses}
From an astrophysical standpoint, nuclear masses can provide
insight into properties of nucleosynthesis.  As an example,
the rapid neutron capture process - or \textit{r process} -
is thought to be responsible for production of roughly
half of all nuclei heavier than iron \cite{iliadis} and
nearly all of the actinides.  Because the r-process proceeds
via sequential neutron captures on existing nuclei, 
$^{A}$Z+n$\rightarrow^{A+1}$Z + $\gamma$, the
path of the r-process is strongly related to the 
mass of r-process progenitor nuclei.  The classical
r process path roughly follows a line of constant neutron
separation energy, $S_n$, where the neutron separation
energy is defined to be the amount of energy necessary
to remove a neutron from a nucleus, $S_n\equiv (M(Z,A) - M(Z,A-1))c^2$.
Knowledge of neutron separation energies is important
in understanding which nuclei are of particular importance
for the r-process.   In thermal equilibrium, the value of
$S_n$ depends on the environmental temperature.  For a
useful assessment of the properties of nuclei along the r
process path, the value of the neutron separation energy
should be known to within $kT\sim$100 keV, corresponding to
the temperature of the r-process environment.

In other astrophysical applications, nuclear masses provide insight
into the thermodynamics of the environment.  Knowing the 
masses of nuclei involved in a reaction will allow for 
a calculation of the energy released in a reaction, which
can provide information on environmental heating.  Examples
where this might be important include neutron star crust heating \cite{meisel15,meisel16}, x-ray burst 
thermal properties \cite{rogers09}, stellar heating, and 
explosive nucleosynthesis.  
\subsection{Example: Nuclear Masses of Neutron-Rich Nuclei and the Astrophysical r Process}
\label{nucmasses}
The astrophysical r process is mentioned here as one example of an astrophysical site in which
nuclear mass measurements may be useful.  In the approximation of
(n,$\gamma$)$\rightleftharpoons$($\gamma$,n) equilibrium, corresponding to a 
the classical r process, nuclear masses can be used for 
a direct determination of neutron separation energies $S_n$.  
Though the canonical r-process may only have a limited contribution
to galactic abundances, nuclear masses are still useful for 
understanding shell structure far from stability \cite{pfeiffer01} for any
r-process scenario.  Minor shell closures, the disappearance of
shells, and the appearance of new shell structure can all inform theoretical 
development and evaluations of r-process sites.

Much of our current understanding of nuclei along the r-process path comes from extrapolating current
mass models to the r-process progenitor nuclei.  Of course, in the
absence of actually measuring r-process progenitor nuclei, pushing studies of nuclear properties to
more neutron-rich nuclei is advantageous.  Thus, there are compelling reasons to study
nuclei with increasing neutron richness as any extrapolation will then become more robust.  These studies may
be applicable to constraining mass models used in the r-process.

The impacts of nuclear mass uncertainties on an understanding of the r process have been studied in detail
in prior work \cite{martin16,mumpower15,atanasov15}.  Shell structure far from stability is important to
understand, but recent work has provided some interesting results which also indicate that 
the uncertainties in masses of the rare-earths must be constrained to better understand production
of these elements \cite{mumpower15,mumpower15b}.
\begin{figure}
	\vspace{-10pt}
	\includegraphics[width=1.0\linewidth,trim={4cm 4cm 4cm 4cm}]{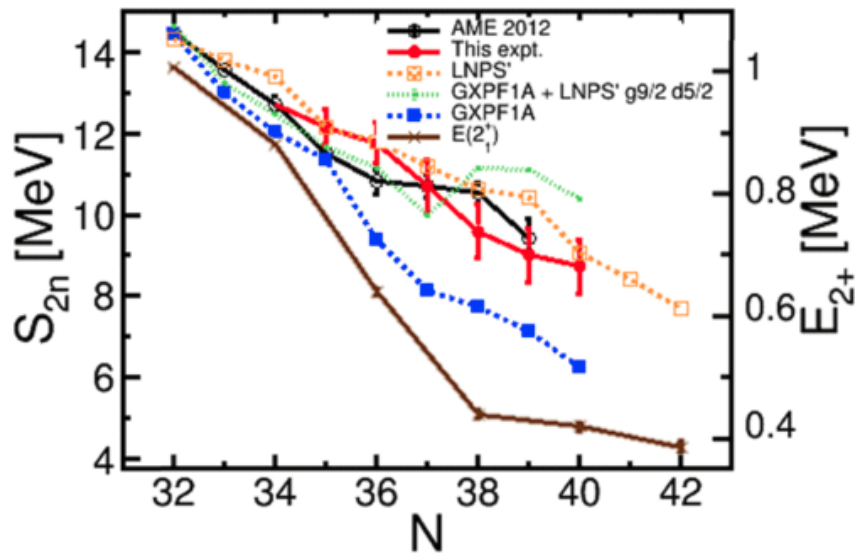}
	\caption{Two-Neutron Separation energies for calcium nuclei (with the energy of the 2$^+$ yrast state indicated) for recent experimental
		work \cite{meisel16} compared to various evaluations.  Used with permission \cite{meisel16}.}
	\label{s2nni}
	\vspace{-10pt}
\end{figure}

Mass models used to determine S$_n$ for
neutron-rich nuclei can disagree significantly.  Some
mass models predict smooth trends in S$_n$ along isotopic lines while others may predict shifts in the slope of S$_n$ as nuclei move further from stability, correlating with nuclear shape changes.  Regions of ``shape coexistence'' may exist in the table of isotopes where nuclei along isotopic lines go from being prolate to oblate (or vice versa).  Of course, different models result in different astrophysical results in various scenarios. 

As an example of the importance of mass measurements, consider Figure~\ref{s2nni}, which shows S$_{2n}$ trends along the Ca isotopic chain for 
experimental work compared to various theoretical models.  The differences become more pronounced with neutron richness.  This is not 
surprising as models can only be compared to and optimized with known masses. Mass measurements are necessary to constrain models for more exotic nuclei. 
Along isotopic chains, even measuring masses that are one nucleon beyond known masses can provide significant constraints on existing models.

Good overviews of nuclear data needs for astrophysical applications can 
also be found in Refs. \cite{kajino19,sun15,matos09}, and \cite{rahaman08}.
\section{Techniques}
\begin{figure}
	\includegraphics[width=\textwidth]{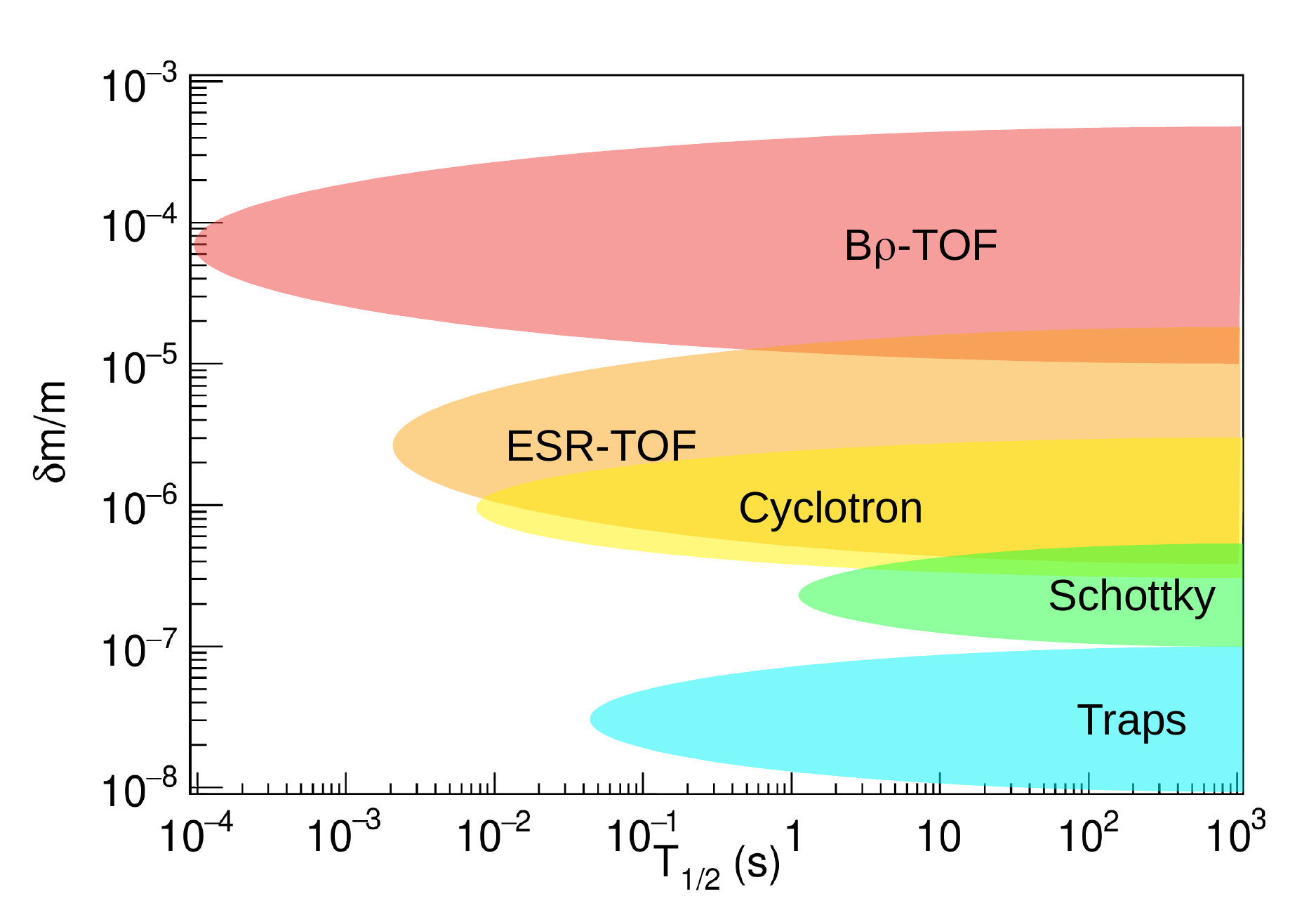}
	\caption{\label{mass_params}Approximate resolutions and lifetimes of various mass measurement techniques \cite{kajino19}.  Used with permission
		\cite{kajino19}.}
\end{figure}

Techniques for measuring nuclear masses include the use of Penning traps, beamline
TOF techniques, and storage rings.  While numerous mass measurement techniques
are available \cite{mittig93}, we concentrate on those currently used for mass measurements of
exotic nuclei.  Each technique has its own advantages and 
disadvantages.  Mass measurement techniques are limited by the half-lives of nuclei
they are capable of measuring.  Likewise, each technique has a mass resolution limit.
The approximate range of lifetimes and mass resolutions accessible by various techniques is 
shown in Figure \ref{mass_params}.

Mass measurements of exotic nuclei with radioactive ion beams 
(RIBs)
come with constraints ubiquitous to the fact that 
unstable beams are being used.   Using RIBs to measure masses
means that any method used must be able to measure 
potentially short-lived nuclei, and these methods are
also constrained by any rate limitations of the facility that 
produces these nuclei.  Methods adopted to measure masses 
are specialized to deal with one or both of these constraints.

Direct measurements of nuclei include frequency-based and time-based measurement \cite{sun15}.  Of the frequency-based measurements, Penning traps and storage rings are the most developed and well-known techniques.  Time-of-flight (TOF) measurements vary in technique, but all rely on measuring the flight time of nuclei of unknown mass between two points in a beam path of known rigidity and comparing to that of nuclei for which the masses are known. 
\begin{figure}[h]
	\includegraphics[width=0.9\textwidth, trim = {0 0cm 0 1 cm}]{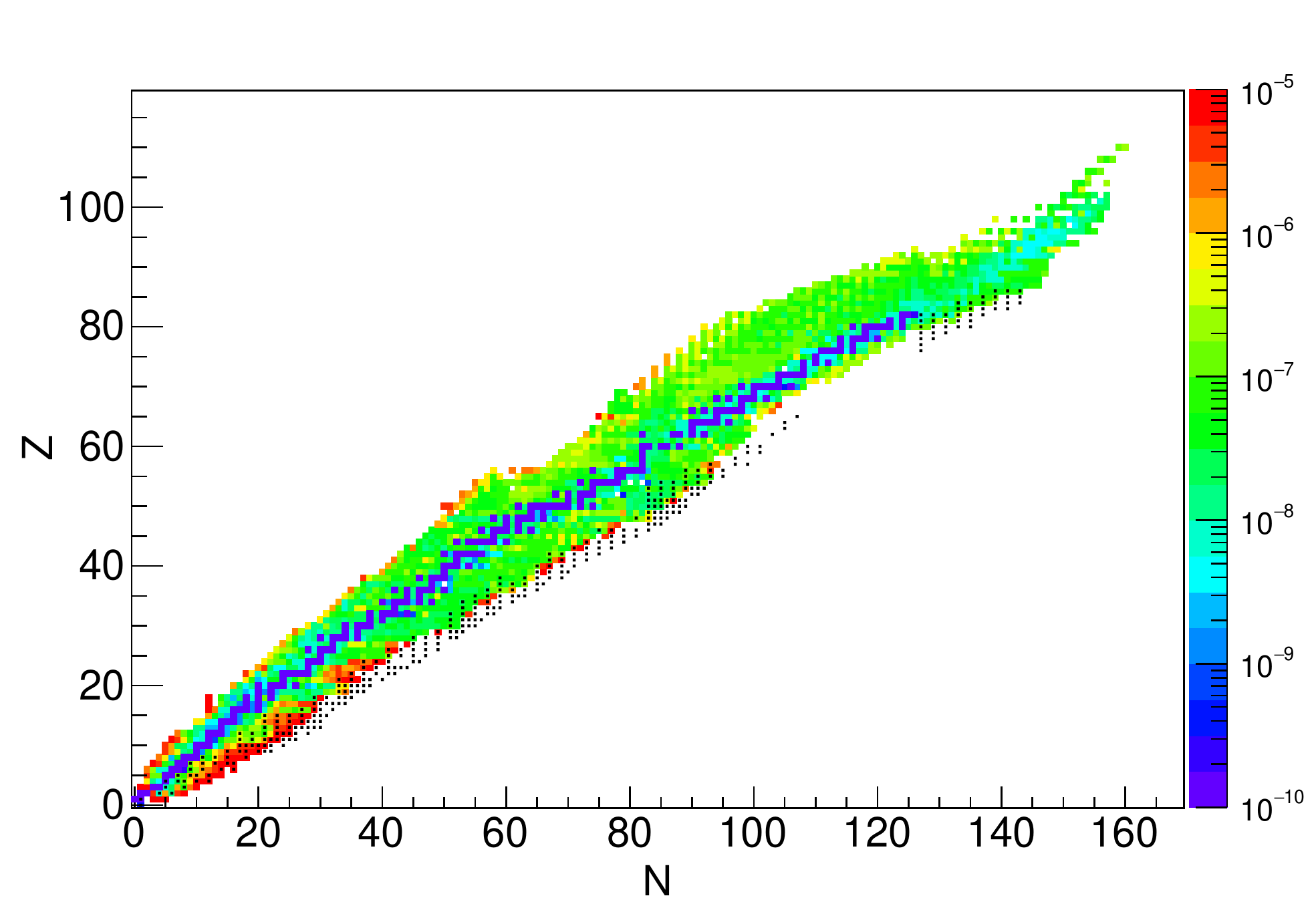}
	\includegraphics[width=0.9\textwidth, trim = {0 0cm 0 1 cm}]{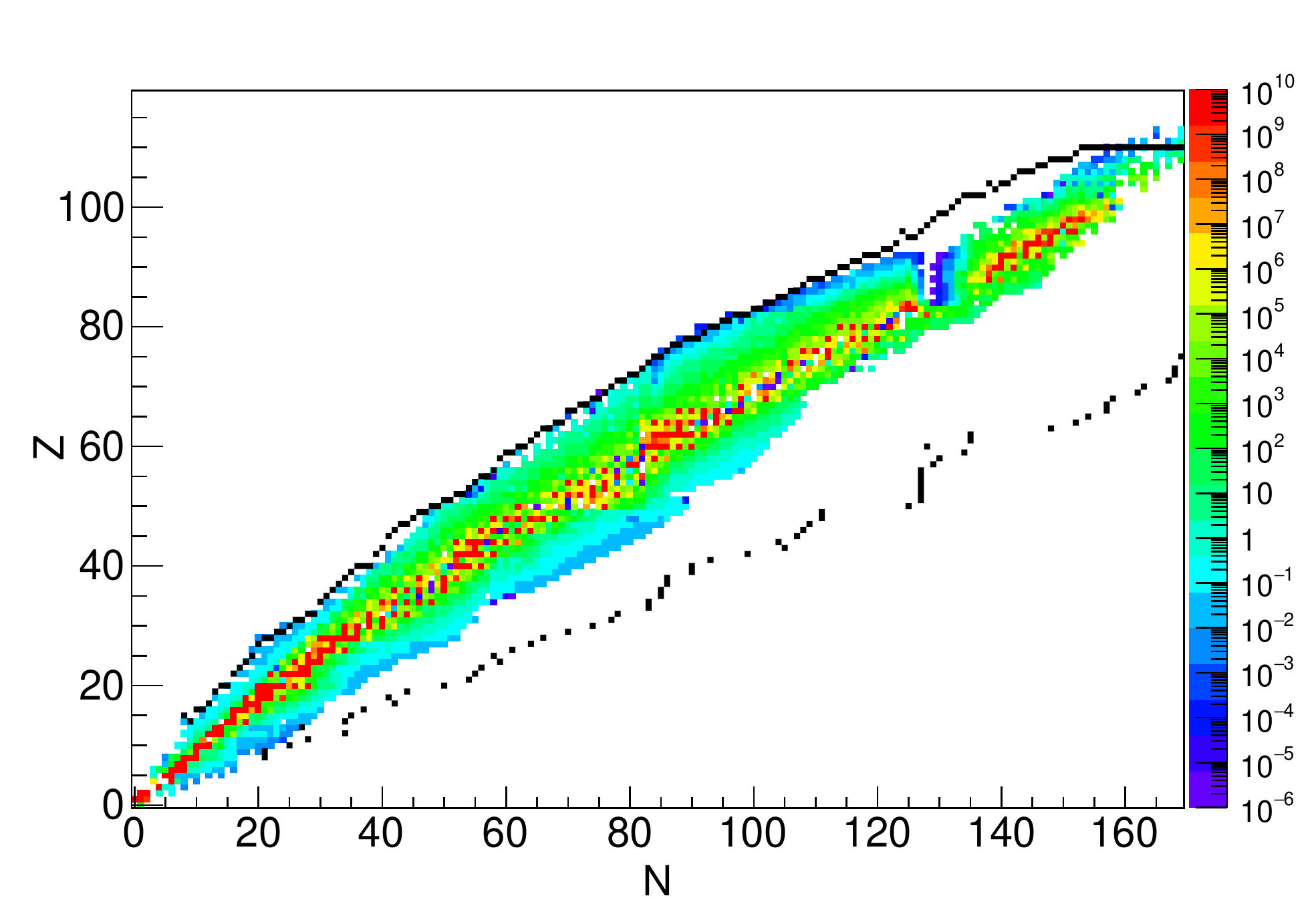}
	\caption{\label{mass_unc}Top: Relative uncertainties $\delta$m/m of nuclear measured nuclear masses \cite{ame16a,ame16b}. 
		Bottom: Measured decay rates of nuclei (s$^{-1}$).  The approximate location of the r-process path is shown in 
		the top figure, and the approximate locations of the drip lines are shown on the bottom \cite{kajino19}. 
		Used with permission \cite{kajino19}.}
\end{figure}

Given the number of techniques and efforts to measure nuclear masses 
with greater precision, constraints on nuclear masses and mass models for nuclei approaching the r-process path has seen significant progress in the past decade.  A representation of the relative uncertainty in nuclear mass measurements as of the AME2016 \cite{ame16a,ame16b} update is shown in Figure \ref{mass_unc}. While the masses of most r-process nuclei have yet to be studied, progress
continues to be made.
\subsection{Penning Traps}
\begin{figure}[h]
	\includegraphics[width=\textwidth]{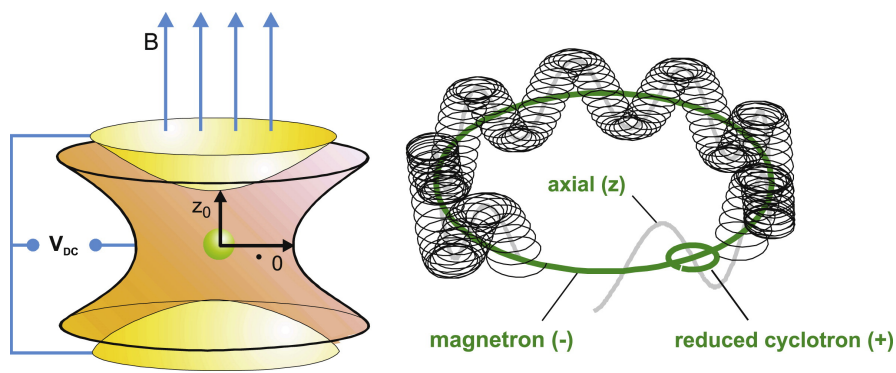}
	\caption{\label{penning_trap}Left: Schematic of a Penning trap spectrometer showing the electric field plates and the magnetic field.  Right: Example of the path of an ion in a Penning trap \cite{ringle09} showing the three motions associated with the eigenfrequencies described in the text.  Used with permission \cite{ringle09}.}
\end{figure}

Perhaps one of the more precise methods for nuclear mass measurements is the Penning trap.  This device
functions by creating a longitudinal uniform magnetic field in a quadrupole electric field.  A schematic 
a Penning trap is shown in Figure \ref{penning_trap} \cite{ringle09}.  Penning traps have been
used to measure atomic and nuclear masses with resolutions of $\delta m/m \lesssim 10^{-8}$ for nuclei
with half-lives as low as a few ms \cite{dilling18}.  They are also useful for mass measurements of low intensity beams.  (For excellent reviews of Penning trap mass spectrometry, see Refs. \cite{dilling18} and \cite{brown86}.)

In Figure \ref{penning_trap}, if the magnetic field defines the $\hat{z}$ axis in cylindrical coordinates,
$\mathbf{B} = B_\circ\hat{z}$, then a quadrupole electric potential is defined by \cite{dilling18}:
\begin{equation}
\phi(r,z) = \frac{\phi_\circ}{4d^2}\left(2z^2 - r^2\right)
\end{equation}
where $d$ is a dimension characterizing the trap.  The magnetic field in this configuration confines
ions in a plane perpendicular to $\hat{z}$.  The radial motion of the ions in the trap are defined by
this field.  The quadrupole electric potential confines ions axially.  The cyclotron motion of ions in the trap is then
related to the magnetic field, where the cyclotron angular frequency, $\omega_c$, is:
\begin{equation}
\omega_c = \frac{qB}{m}
\end{equation}  
In addition to the cyclotron oscillation, the ion can oscillate axially as a result of the electric field.
Its motion can be shown to be simple harmonic with a frequency \cite{brown86}:
\begin{equation}
\omega_z^2 = \frac{q\phi_\circ}{md^2}
\end{equation}  
The cyclotron frequency in the absence of the external electric field is shown above.  However,
with the external electric potential, the equations of motion are solved to result in three
independent eigenfrequencies.  One of these is the axial oscillation $\omega_z$. Another is a modification of
the cyclotron motion $\omega_+$, and a third is an oscillation due to $\mathbf{E}\times\mathbf{B}$ drift
motion.  The angular frequencies $\omega_\pm$ are \cite{brown86}:
\begin{equation}
\omega_\pm = \frac{\omega_c}{2}\pm\sqrt{\left(\frac{\omega_c}{2}\right)^2 - \left(\frac{\omega_z}{\sqrt{2}}\right)^2}
\end{equation}
where $\omega_+$ is known as the modified or reduced cyclotron frequency and $\omega_-$ is known as the magnetron frequency.  It's instructive to note that $\omega_c>\omega_+\gg \omega_z>\omega_-$ \cite{dilling18} and that $\omega_c = \omega_++\omega_-$.
The eigenmotions defined by $\omega_z$ and $\omega_\pm$ are shown in Figure \ref{penning_trap}.

An example of a typical mass measurement with a Penning trap is described 
in Refs. \cite{ringle09} and \cite{dilling18}. A typical measurement using a Penning trap would involve first cooling the ions to energies of a few 
keV using degraders.  Gas cells are commonly used as degraders owing to their excellent uniformity and adjustability.  
The ions are then electrostatically guided to the trap,
where their motion can be excited through resonance frequencies with RF fields.  At the resonant
cyclotron frequency, the ions are excited and then extracted from the trap, where their energy can
be measured via time-of-flight from the trap to a TOF detector.  The TOF measurement need not be
as precise as the resonance frequency measurement, as one is only looking for the frequency at which the 
ion energy results in a resonance due to the cyclotron resonance being set up (and, hence, 
the TOF at resonance is dramatically different that off resonance).  A
resolution of a few ns is sufficient \cite{ringle09}.  

The time it takes to complete this process of cooling, exciting, extracting, and measuring TOF is the major limitation of mass 
measurements using Penning traps.  Such measurements typically consume a few 10s to 100s of ms, placing 
a lifetime limit on the isotopes that can be measured.  This tends to limit the range of 
isotopic species that can be measured.  However, the excellent precision is extraordinarily useful as
Penning trap mass measurements can be used as calibration points for possibly lower-resolution techniques that can effectively measure masses of more exotic species.

A number of trap facilities worldwide have measured both proton-rich and neutron-rich nuclei with
excellent accuracy.  Because of the lifetime limitations of traps, extremely exotic species - such
as those on the r-process path - have fewer measurements.  However, where measurements exist, the
resolution is sufficient to use these as calibration points for less-precise methods that can
measure the masses of the more exotic species.  

\begin{figure}
	\includegraphics[width=\textwidth]{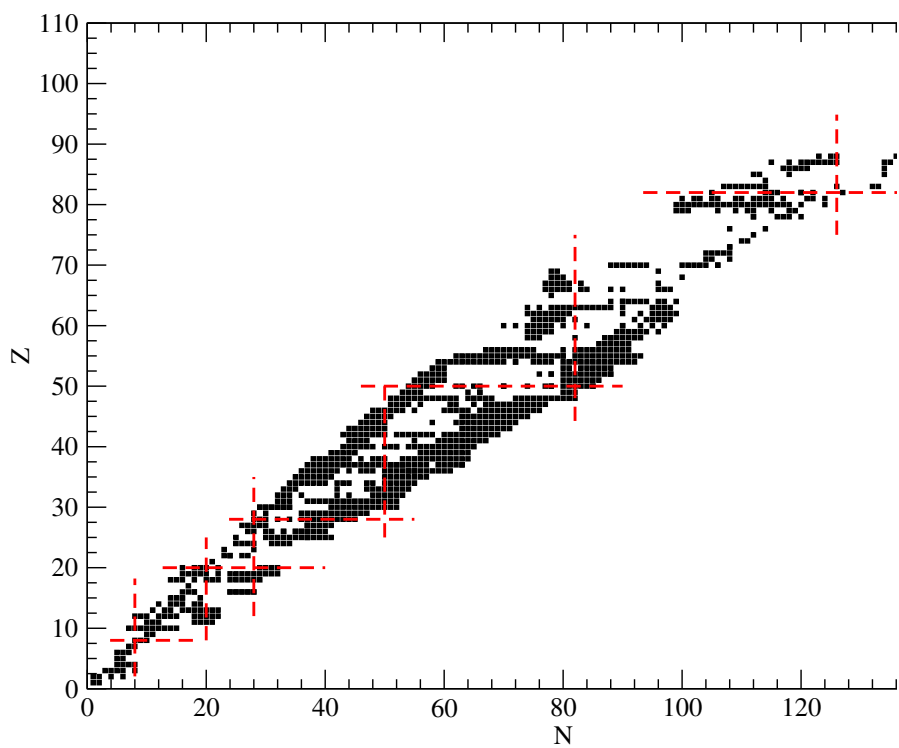}
	\caption{\label{trap_measurements}Isotopic species that have been measured with Penning traps \cite{dilling18}.  Isomers and charge states
		are not distinguished in this figure.  Nuclear shell closures are indicated by the dashed lines.}
\end{figure}

Figure \ref{trap_measurements} shows the masses of various nuclei that have been measured
with Penning traps.  A brief summary of a representative sample of Penning trap devices is given below.
\subsubsection{CPT} 
The Argonne National Laboratory Canadian Penning Trap (ANL-CPT) has been inserted as a functional component of
the ATLAS linac \cite{savard01,wang04,schelt12}.  Significant progress has been made with the CPT to measure 
the masses of isotopes along the N=Z line.  However, measurements have also been made of the masses of
nuclei approaching the r-process path \cite{schelt12}.  This device is capable of mass accuracies of $\delta m/m\sim 10^{-8}$.

In addition to the Penning trap, the CPT is coupled to a radio-frequency quadrupole (RFQ) trap.  Species are produced 
in the ATLAS facility and analyzed in the ANL Enge split-pole spectrograph. Prior to trap injection, nuclei are cooled in
a gas cooler/buncher system.

The recent CARIBU upgrade at ANL has increased the reach of nuclei available at the CPT \cite{schelt12}.  Recently, 
40 neutron-rich nuclei with $51\le Z\le 64$ have been measured, along with Te and Sb nuclei along the r-process path \cite{schelt12}.
\subsubsection{ISOLTRAP} 
Over 400 nuclear masses have been measured with the ISOLTRAP tandem Penning trap facility located at the
online isotope separator ISOLDE at CERN.  Nuclei with lifetimes as low as $\sim$50 ms have been 
measured with typical accuracies of $\delta m/m\sim 5\times 10^{-8}$ \cite{isoltrap,wolf13}.

The ISOLTRAP facility is enhanced with the addition of a multi-reflection
time-of-flight (MR-TOF) mass separator and spectrometer \cite{wolf13}.  (See
\S\ref{other_methods} for a brief description of the MR-TOF method.)
The primary purpose of this is to provide an additional level of beam purification.  Because
of
the exceptional beam purity provided, the ISOLTRAP facility is also used to measure
half-lives of exotic nuclei in a background-free environment \cite{wolf16}.  The half-life
of $^{27}$Na was recently measured with this setup \cite{wolf16}.  

While the list of nuclei measured with ISOLTRAP is extensive, notable cases include the masses of
nuclei between between $^{17}$Ne and $^{232}$Ra \cite{herfurth05,baruah08}.
In addition, ISOLTRAP has been used to measure the masses of nuclei near the $N=82$ shell closure.  Current
results are also pushing the limits of precise neutron-rich masses past the $Z=82$ shell closure.  This
is challenging given the lifetime limitations off Penning traps and the lifetimes of r-process
progenitor nuclei.
\subsubsection{JYFLTRAP} 
JYFLTRAP is located at the IGISOL facility at the University of
Jyv\"{a}skyl\"{a}\cite{hakala12,aysto13,kankainen11,rahaman07}.
It has achieved mass accuracies near 50 keV and generally produces results
with $\delta m/m\sim 10^{-8}$ \cite{jyfltrap,jyfltrap2}. To date, it has
measured nearly 400 masses up to A$\approx$120.  Of these, roughly 300 are 
masses of neutron-rich nuclei \cite{rahaman07a,hakala08,Rahaman2007b,hager06,hager07,hager07b,hakala11,hakala12,wieslander09,rinter07}.
These include masses near the Z=50 r-process waiting point as well as many of the
neutron-rich rare earth nuclei.
\subsubsection{LEBIT} 
The Low Energy Beam and Ion Trap (LEBIT) facility has been a leader in trap facilities in the USA.
The LEBIT trap takes advantage of the radioactive ion beam production capabilities of the National
Superconducting Cyclotron Laboratory by slowing the fragmented fast beams prior to trap 
injection \cite{ringle09,ringle13}. Fragments produced in the NSCL coupled-cyclotron facility (CCF) and
separated in the A1900 fragment separator are slowed in the NSCL gas stopping system \cite{ringle09}.
The process of combining the analysis beam, gas stopping system, and trap is essential for utilizing 
a trap in a facility that provides a large variety of fragmentation beams.  

Slowing and stopping of the beams in LEBIT is accomplished with a linear gas stopper and cyclotron gas
stopper \cite{schwartz16}.  Prior to injection into the Penning trap, a beam accumulator and buncher cools the beam \cite{schwartz16b}. 

Many of the masses which are difficult or impossible to produce at ISOL facilities can be produced at the NSCL, making LEBIT complementary to existing facilities.  As a result, the
LEBIT facility has measured the masses of over 50 isotopes with high precision, with a concentration on 
the proton-rich nuclei.  As the NSCL transitions to the Facility for Rare Isotope Beams (FRIB), new opportunities may be available for LEBIT \cite{gade16}.   
\subsubsection{SHIPTRAP} 
The GSI SHIPTRAP facility has concentrated primarily on proton-rich nuclei.  It has also excelled at measuring the masses of more massive nuclei 
\cite{herfurth05}.  Between ISOLTRAP, SHIPTRAP, and JYFLTRAP, devices on the European 
continent have measured a very large range of masses with the Penning trap technique.  It consists of a two-stage trap system
with the intent of searching for super-heavy elements. It is thus optimized for the transmission and measurements of transuranic nuclei with $Z\ge$92, filling a niche among the trap facilities \cite{schonfelder02}.

Early measurements with SHIPTRAP include $^{147}$Ho$^+$,  $^{147}$Er$^+$, and $^{148}$Er$^+$ \cite{block05} with latter measurements 
prepared to examine the endpoint of the rp-process \cite{block06}.
\subsubsection{TITAN} 
The TITAN facility (TRIUMF's Ion Trap for Atomic and Nuclear science) consists of several Penning traps.  
Recently, TITAN underwent an upgrade to install a charge breeder in its electron beam ion trap \cite{lascar16}.  This 
allows for the production of higher charge states, resulting in improved mass resolution.  

For decay spectroscopy, the facility utilizes photon counters and seven Si(Li) detectors for simultaneous in-trap decay
spectroscopy experiments \cite{dilling06,delheij06,froese06,kwiatkowski13,lascar16}.  TITAN has been used
for nuclear mass measurements and mass model constraints \cite{dilling03}, though its primary purpose is
determinations of the $V_{ud}$ CKM matrix elements.  Precise measurements of the masses
of parent and daughter nuclei are necessary for precise determinations of $V_{ud}$.

Recently, TITAN has been used to measure the neutron-rich nuclei $^{98,99}$Rb and $^{98-100}$Sr \cite{klawitter16}.
This is useful for nuclei approaching the r-process path.  In addition, the masses of nuclei approaching the N=82 closed shell, $^{125-127}$Cd have also been measured \cite{lascar17}.  A notable feature of TITAN is its ability to measure nuclear masses with very low 
half-lives, such as the mass of $^{11}$Li,
with $t_{1/2}$ = 8.8 ms \cite{smith08}.
\subsection{Other Trap Facilities}
While the Penning trap itself is small and relatively straightforward to manufacture, the equipment 
necessary to successfully operate the trap, including the field generators, beam coolers, power supplies,
and RF sources makes the operations of the trap precise and the requirements exacting.

However, multiple traps have been installed about the world in addition to those described above.
Other successful Penning trap programs and those planned or under construction include the Florida State University trap \cite{fsu_trap},
SMILETRAP \cite{smiletrap}, MATS \cite{mats}, PIPERADE \cite{piperade}, WITCHTRAP \cite{witchtrap},
REXTRAP \cite{rextrap}, TRIGA-TRAP \cite{trigatrap}, the Mainz trap \cite{mainztrap},
HITRAP \cite{hitrap}, PENTATRAP \cite{pentatrap}, MLLTRAP \cite{mlltrap}, the Lanzhou trap \cite{lanzhou_trap},
and the RIKEN trap \cite{riken_trap}.  These are dedicated to various studies involving the
measurements of nuclear masses as well as fundamental properties of matter and anti-matter.
\subsection{Time-of-Flight Methods}
As the name implies, time-of-flight (TOF) methods are used to determine nuclear masses 
by measuring the flight time of nuclei between two points in an spectrometer.
The spectrometer is calibrated with the masses of known nuclei.  Thus, the 
accuracy of mass measurements with TOF methods depends on the uncertainties of
the calibration masses, though these uncertainties are not necessarily the largest contributor
to the total measurement uncertainty.

Mass measurements determined via TOF are generally less accurate than those determined with traps.  However, while traps
are limited by the lifetime of the measured nuclei, 
the lifetime of nuclei measured with TOF method is only limited
by the time for nuclei to traverse the flight path of the spectrometer used.  This is generally well under 1 $\mu$s.

Beamlines used in the TOF method can consist of a ``tagging section'' in which flight time is measured between two
distinct points in a spectrometer.  Alternatively,
closed-loop flight paths (rings) allow for the particle
of interest to pass the same point in the beamline multiple times.  The ring method is generally more accurate as
measurements can be averaged over multiple particle passes,
and systematic error can be reduced as only a single TOF 
detector is necessary.
\subsubsection{Rigity-Based TOF Measurements}
The B$\rho$-TOF method is used to measure nuclear masses through the measurement of the flight-time
of nuclei between two points in a beam facility \cite{matos09}; thus, it utilizes a tagging section of
a spectrometer.  The tagging section is typically
$\sim$100 m in length in facilities employing this method.  
This experimental
setup is conceptually straightforward and the
associated device parameters can be varied significantly to 
suit the goals of a experiment. However, though this method
employs detectors typically used in a variety of nuclear physics
experiments, they must be operated and understood very precisely.

In many measurements employing the B$\rho$-TOF method,
beamline diagnostics are used (or substituted with devices of
higher precision).
In a typical setup, two or more detectors are placed
at the ends of the beamline tagging section to 
determine the flight time of short-lived species.
The TOF measurement is complemented with
a measurement of the particle rigidity $B\rho$, thus the
reason for referring to this method as the $B\rho$-TOF
method.

Radioactive ion beam accelerators are, to first order, 
momentum selectors.  In the simplest approximation particle momentum is
related to the magnetic field and turn radius in
the bend of a beam path through the field.  
The relationship between particle mass and rigidity can be 
easily derived:
\begin{equation}
\frac{B\rho}{\gamma} = \frac{mv}{q}
\end{equation}
where $\gamma$ is the usual Lorentz factor for particles in the beamline.  
With an overall TOF uncertainty of $\sim$50 ps (after correcting for path length via a measurement of
the particle rigidity), and a TOF $\sim$ 0.5 $\mu$s, the uncertainty in 
the mass from the TOF is $\sim$10$^{-4}$.  Greater statistics can reduce the 
average
statistical uncertainty to values typically $\sim$10$^{-6}$ to $\sim$10$^{-5}$.
This corresponds to a mass resolution of $\sim$100 keV/c$^2$, which is roughly the temperature,
$kT$,
of the r-process environment.  This resolution is generally sufficient to constrain
mass models far from stability.

The lifetime limitation
of measured nuclei is due to the TOF between two detectors.  Calibration of 
flight times with this method is accomplished by incorporating known ``reference masses'' in 
the mixed beam.  Because the B$\rho$-TOF method can measure shorter lifetimes, it can 
extend the range of known masses to more neutron-rich nuclei.  This method is also useful
for measuring multiple masses simultaneously in a single experiment. 

Reference masses for calibration of
the method can be taken from masses measured with Penning traps.  The major limitation in the limit
of nuclei that can be measured is then the production rates of exotic beams.  

Particle rigidity with this method can be determined by measuring beam position at
dispersive planes in the beamline.   Additional complications of this method may be the 
incorporation of charge states (species that are not fully ionized or species in
multiple ionization states) in the beam, which must be
separated by measuring the energy deposition in a particle telescope at the end of the beamline.  This method is sometimes referred to as the $B\rho$-TOF-$\Delta E$ method.

A representative sample of mass measurements at facilities incorporating the $B\rho$-TOF or $B\rho$-TOF-$\Delta E$ 
method are described below. 
\subsubsection{NSCL} 
\label{nscl_desc}
\begin{figure}
	\includegraphics[width=\textwidth]{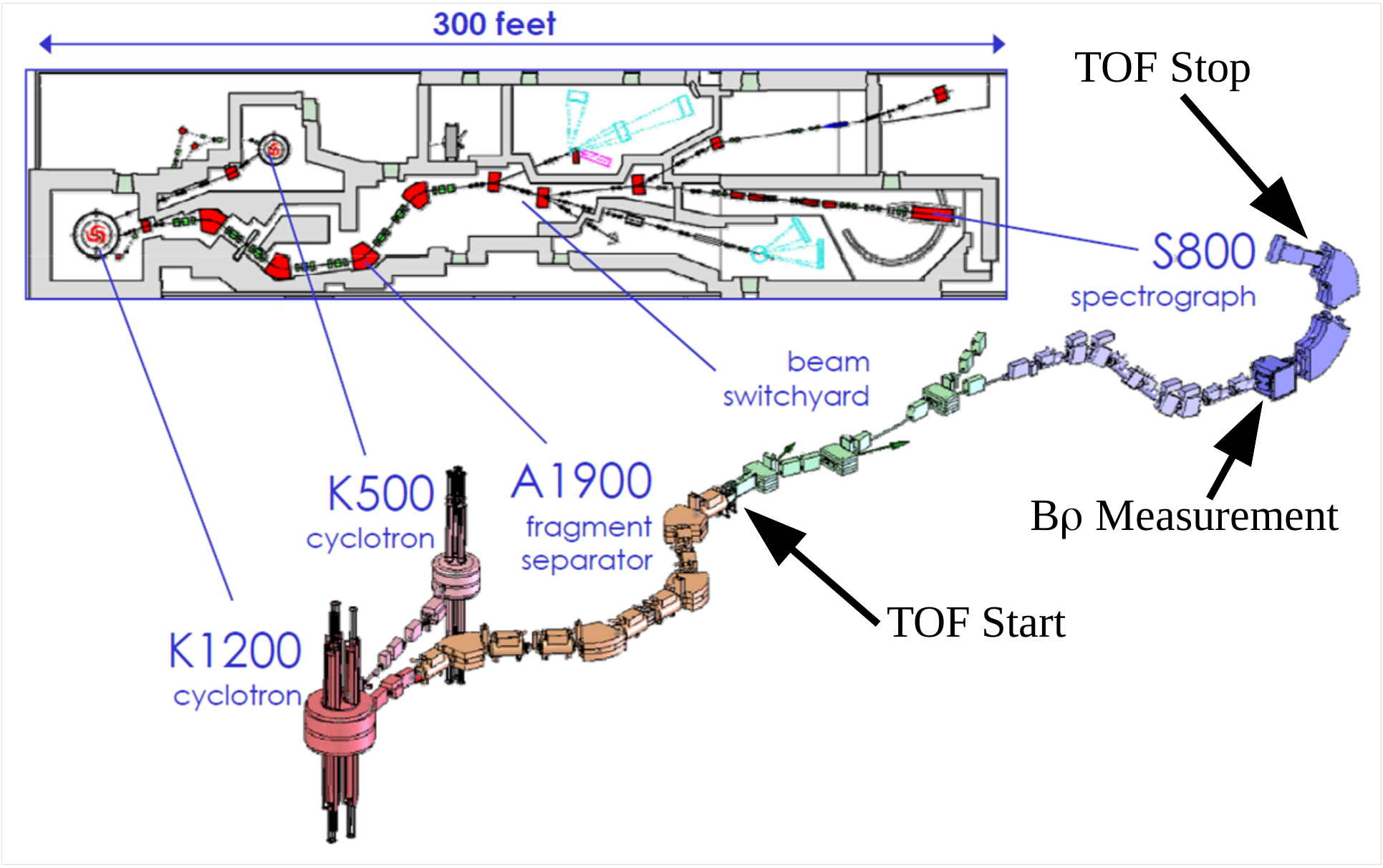}
	\caption{\label{tof_nscl}Schematic of the TOF-B$\rho$ experimental setup at the NSCL facility \cite{estrade_tof}.}
\end{figure}

Located in East Lansing, MI, the National Superconducting Cyclotron Laboratory (NSCL) in East Lansing, MI 
currently utilizes the coupled-cyclotron facility (CCF) and the A1900 spectrometer coupled to the 
S800 spectrometer to create a particle flight path of about 88 m.  A schematic of a typical NSCL TOF mass measurement experiment is shown in Figure \ref{tof_nscl}. 

The TOF detectors
are placed at the A1900 extended focal plane and the focal plane of the S800.   Particle 
rigidity is measured at the target position of the S800  \cite{estrade07,matos08}.  (See Fig. \ref{tof_nscl}.)
Figure \ref{tof_nscl}.  For such a setup, the tagging section begins at the end of the A1900 spectrograph \cite{a1900} and ends 
at the end of the S800 spectrograph \cite{s800}.  Both spectrographs
are matched in rigidity to allow the particles of interest to 
pass with a minimum loss of beam intensity. 
A point of maximum momentum dispersion is located in the S800, usually at the target location.  While the S800 can
be operated with a dispersive plane at the end station,
running in the mode with the dispersive plane at the target location and a focus at the S800 terminus allow for the use
of small TOF timing scintillators in the device, which
can reduce the uncertainty and subsequent necessary corrections
to particle position in the scintillators. 

In order to measure the particle position at the S800 focus, the detector used must have as little material as possible
interfering with the beam to reduce the amount of energy straggling in the beam, which adds an additional uncertainty.
Ideally, any detector used should be placed as close to the
end of the tagging section as possible.  The reason for this
is because the uncertainty induced in the velocity, $\delta v$ will result
in an uncertainty in the $\delta$ TOF of the beam particles.  If the
length between the rigidity measurement and the end of the tagging section, $d$, is shorter, the shift in 
TOF is then $\delta$ TOF $= d/\delta v$.  
A detector of choice 
for such a task would be a micro-channel plate (MCP) based on secondary electron emission 
\cite{shapira00,rogers15}.  Such devices can introduce $\sim$50 $\mu$g/cm$^{2}$ of
material into the beamline while providing very fast ($\lesssim$ 1 ns) time resolution and excellent 
position resolution ($\sigma_x\sim 500~\mu$m).

\begin{figure}
	\includegraphics[width=\textwidth]{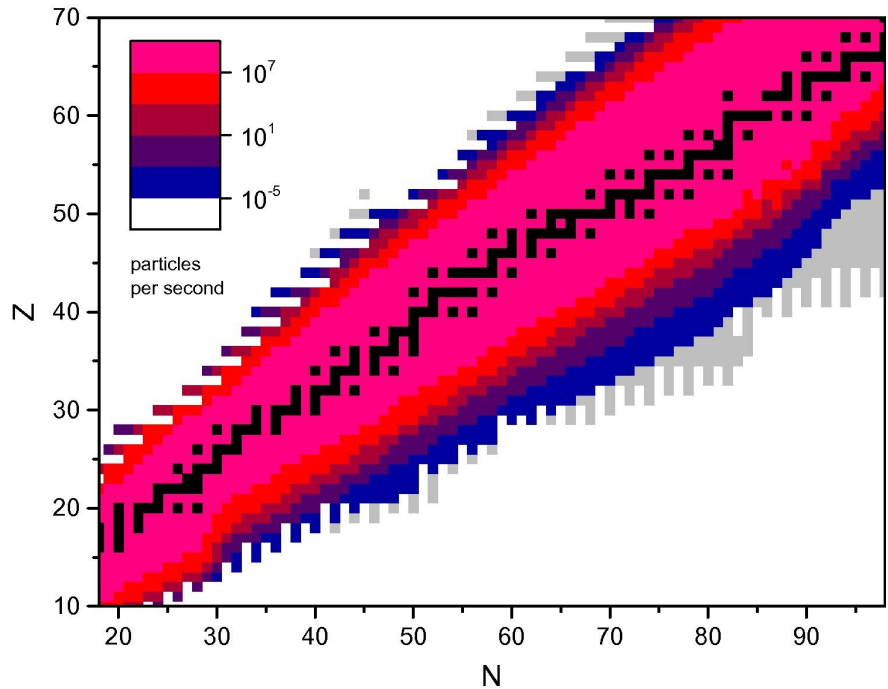}
	\caption{\label{frib_rates}Predicted isotopic production
		rates at the Facility for Rare Isotope Beams \cite{FRIB}. 
		Used with permission \cite{FRIB}.}
\end{figure}

By the early half of the decade starting in 2020, the NSCL will transition to the Facility for Rare 
Isotope Beams (FRIB).
As a result the production rates of more exotic nuclei will
make it possible to feasibly measure nuclear masses farther
from stability using the B$\rho$-TOF technique.
Predicted production rates for nuclei at the FRIB are
shown in Figure \ref{frib_rates}\cite{FRIB}.  It is predicted that 
nuclear masses for nuclei well past the second r-process 
peak into the rare earth region will be possible \cite{HRS_WP}.  In
addition, the rigidity resolutions afforded by the planned High Resolution
Spectrometer (HRS), should allow for sensitive rigidity corrections for
TOF measurements at the HRS \cite{baumann16}.

Over the past several years, the NSCL has developed several mass measurement
experimental programs \cite{schatz13}.  In particular, neutron-rich
nuclei of increasing mass have been a target of many recent experiments 
\cite{meisel16}.  Recent published results include the measurements of
$^{59-64}$Cr and neutron-rich Ar, Sc, Mn, and Fe relevant to crustal heating
in neutron stars \cite{meisel15,meisel16}.  At the time of this writing,
analysis continues on mass measurements to evaluate the shape coexistence of
nuclei near the r-process path during the early onset of the r-process.
Additional measurements using this technique at the NSCL have determined 
the masses of $^{61}$V, $^{63}$Cr, $^{66}$Mn, and $^{74}$Ni\cite{estrade11,matos12}.  Resolutions achieved were 10$^{-4}$.
\subsubsection{Spectrom\`etre \`a Perte d'Energie du Ganil (SPEG)} 
Located at GANIL in Caen, France, the high-resolution SPEG
contains  an 82 m flight path coupled to an $\alpha$ spectrometer (so named
because the shape of the spectrometer resembles the Greek ``$\alpha$'').
The mass measurement program associated with the SPEG device has
seen a successful program covering decades \cite{savajols01,savajols01b,gomez05}.

While the NSCL program uses plastic scintillators as the timing detectors of the 
tagging section, the SPEG program uses microchannel plate detectors \cite{shapira00,rogers09} which have an intrinsic time resolution of $\sim$100 ps (FWHM). Two independent rigidity measurements are also
employed; one is at the dispersive focal
plane while the other is made after the TOF tagging section.

Many of the SPEG measurements have concentrated heavily on neutron-rich nuclei
with N=16, 20, and 28.  This may be of particular importance for the ``light
element primary process'' (LEPP).  

SPEG measurements of neutron-rich masses near the N=20 shell closure have 
been used to verify a region of shape coexistence about
N=20 and N=28 nuclei.   These measurements targeted the masses of nuclei with $29\le A \le 47$ \cite{sarazin01}. 
The SPEG programs has also resulted in mass measurements of several light, neutron-rich
nuclei \cite{gillibert86, gillibert87} with resolutions as low as $\sim$10$^{-5}$.
These include measurements of $^{29,30}$Ne, $^{34,35}$Mg, $^{36,37}$Al, and 
$^{31-33}$Na \cite{orr91,sarazin00}.
Additional successful attempts have been made to measure masses of nuclei closer to the
neutron drip line \cite{savajols05, jurado07}, and significant improvements
of neutron-rich masses in the mass region A$\sim$ 10 -- 50 have been made including 
measurements of $^{19}$B, $^{22}$C,
$^{29}$F, $^{31}$Ne, $^{34}$Na, and other light nuclei \cite{gaudefroy12}.  These
measurements were instrumental in confirming the halo structure of several of the light nuclei in this region.  
The C and B nuclei 
are presumed to be at the neutron drip line.  
\subsubsection{Other Facilities}
The SHARAQ spectrometer at the Radioactive Ion Beam Factory in Saitama, Japan
has recently been completed \cite{uesaka08,uesaka12,michimasa13}.
This creates a 105 m TOF tagging sections.  Radiation-hard diamond
detectors are used to sustain high rates with excellent time resolution 
\cite{michimasa13b}.  The SHARAQ spectrometer is very high resolution, allowing
for precise rigidity measurements.  Recently, the masses of $^{55,56}$Ca have
been measured with a resolutions of $\sigma\sim$150 keV and 234 keV respectively.

Another example of a promising facility is the collaborative effort between
Beihang University and the Heavy Ion Reseach Facility in Lanzhou (HIRFL) \cite{xia02}.  Detector development by this collaboration has resulted in fast
plastic scintillators for timing purpose.  These have an intrinsic time
resolution of 5.1 ps \cite{zhao16}.  The HIRFL collaboration has also worked to 
develop high resolution multi-wire proportional counters (MWPCs) with intrinsic
position resolutions of 1 mm for rigidity measurements.  

For light-Z neutron-rich nuclei, the Time-of-Flight Isochronous (TOFI) spectrometer 
\cite{wouters85,vieira86,wouters88} is designed for isochronous mode
operation.  In isochronous mode, the spectometer is set such that all species
have the same flight time while the momentum is used exclusively to determine mass.
Thus, slower-moving ions travel a shorter path, requiring precise position
measurements at the dispersive focal planes.  This measurement is also used 
by rings (described in \S\ref{rings}).  The TOFI spectrometer has been used
to measure the masses of neutron-rich isotopes from Li to P \cite{wouters88,wouters85} with resolutions of 200 to 900 keV.
\subsection{Storage Rings Methods}
\label{rings}
\begin{figure}
	\centering
	\includegraphics[width=\textwidth]{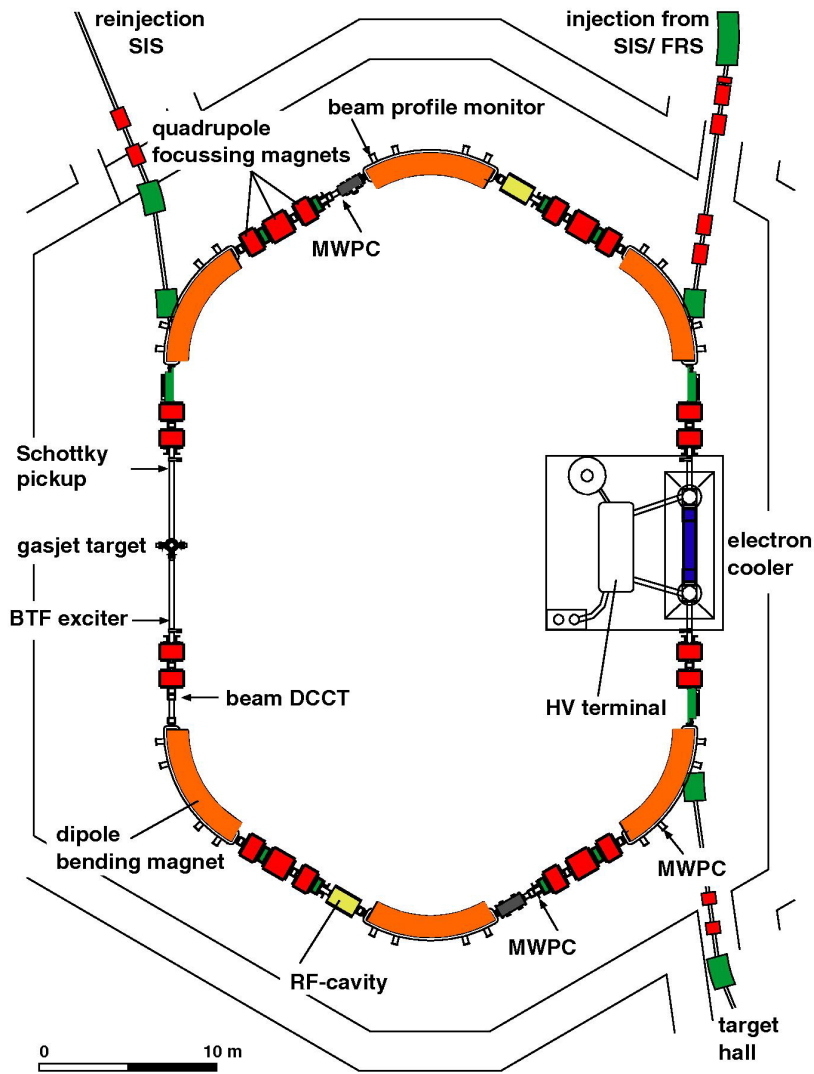}
	\caption{\label{gsi_esr}An example layout of the GSI ESR storage ring \cite{nolden08}. Used with 
		permission \cite{nolden08}.}
\end{figure}

Much like the B$\rho$-TOF method, storage rings operate based on the time of flight of the nuclei being
measured.  However, because the flight path is a closed path, storage rings are capable
of measuring particle passage 
at one point in the ring as the particles pass that point \cite{ozawa12,ozawa14} (though
measurements at more than one point can be made as well).  This method has several
advantages.
One advantage is that only a single timing detector is necessary in the particle flight path.  Of course,
the energy loss of the particle in the detector must be kept to a minimum, but systematic errors
induced by multiple particles can be reduced.
A second very powerful advantage is that the storage ring  is able to measure the a particle several times
as it traverses the loop multiple times prior to decay.  An example of a storage ring is illustrated in Figure \ref{gsi_esr}\cite{nolden08} showing the GSI Experimental Storage Ring (ESR).  Here, multiple components of the ring are indicated, including 
the injection mechanism and measurement locations.

However, unlike the B$\rho$-TOF method, because they make multiple measurements of the same particle, storage rings are typically classified as frequency-based measurements in that they make
multiple measurements of a particle's passage about the ring in a time period.   It has been
shown that the relationship between the number of circuits that a particle traverses about the ring is
related to its mass by the measurement of revolution frequency of ions in a ring of circumference C \cite{nolden08,franzke08}:
\begin{equation}
f = \frac{\beta c}{C}
\end{equation}
For ions in the storage ring, the deviation in individual frequency measurements, $\delta f$, is:
\begin{align}
\label{ring_freq}
\frac{\delta f}{f} &= \frac{\delta \beta}{\beta} - \frac{\delta C}{C}\\\nonumber
~& =\frac{1}{\gamma^2}\frac{\delta\beta\gamma}{\beta\gamma} - \alpha_p\frac{\delta p}{p}\\\nonumber
~&= 
\left(\frac{1}{\gamma^2} - \alpha_p\right)\gamma^2\frac{\delta\beta}{\beta}-\alpha_p\frac{\delta(m/q)}{(m/q)}
\end{align}
with $\gamma$ the Lorentz factor.  The term $\alpha_p$ is known as the ``compaction factor'' which is 
the ratio between the relative orbital path length and the relative momentum $\delta L/L = \alpha_p \delta p/p$ and results from the magnetic parameters in the storage ring.  The third line above results from inserting $p=m\beta\gamma c$ into the second line and expanding.  

From Equation \ref{ring_freq}, it can be seen that the frequency width of a single species is related to the
mass uncertainties in a single species.  Thus, more accurate frequency measurements result in more
accurate mass measurements.  

The resolution of a storage ring can be improved in two ways.  The
first
is by tuning the beam parameters such that $\alpha_p = \gamma^{-2}$, so that the first term vanishes, 
known as ``isochronous mode'' in which the revolution frequency is constant for a nuclear species 
\cite{hausmann00,wollnik97}.  This method
works well for short-lived species, but has the drawback that it requires beam dispersions which can be
large at even moderately relativistic settings.  

Another way is to reduce the spread in beam velocity $\delta\beta/\beta$ through ``cooling,'' effectively
reducing $\delta\beta$ in Equation \ref{ring_freq}. Here, the phase space of the beam is greatly
constrained through a number of methods \cite{caspers13}.
The frequency
of particle revolutions for multiple particles in the ring can then be analyzed as a 
Schottky spectrum, in which all species have the same
average velocity, and shifts in frequency correlate with shifts in mass \cite{schlitt97}.  Cooling is particularly necessary for longer-lived beams when
the momentum dispersion increases due to collisions with other beam particles, beamline gas, space-charge effects, and beam target collisions. One example is that employed by the GSI-ESR and others in which a beam of co-moving electrons is injected into the primary particle beam \cite{franzke87,mao16,lin16,bryzgunov18}.  Coulomb
interactions between the electrons and the beam result in a thermal equilibrium prior to the 
removal of the electrons from the beam, having absorbed some of the transverse momentum of the beam.

Because rings effectively count the revolutions of ions, they are also able to provide relatively 
accurate
measurements of lifetimes by observing the revolution number at which the ion no longer exists 
\cite{litvinov08}.
The lifetime of a single particle is then the number of revolutions a particle makes in the ring multiplied by the time
per revolution.  The measured lifetime then has a resolution that equal to the time for 
one revolution, $1/f$.   
With modern rings, this is $\sim 2$ $\mu$s.

A representative sample of storage ring facilities used for mass measurements is described in what follows.
\subsubsection{GSI-ESR}  
The experimental storage ring located at GSI (GSI-ESR), which was described briefly previously
and in Figure \ref{gsi_esr} \cite{nolden08,geissel08} began operation in 1990.  Particles analyzed
in the ring are produced in the fragment separator (FRS) prior to injection into the ESR.

With a circumference of 108 m and a maximum magnetic rigidity of 10 Tm, the ESR can transmit
U nuclei with a maximum energy of 560 MeV/u at a rotation frequency exceeding 2$\times$10$^6$ s$^{-1}$
\cite{geissel08,kajino19}.

The ESR is capable of running in both Schottky mass spectrometry (SMS) (all particles ideally have the same velocity) or isochronous mass spectrometry (IMS) mode (all particles have the same frequency), both of which
are described above.  In SMS mode, particles in the beam are cooled to the central velocity of the beam,
such that $\delta v/v\rightarrow 0$.

The ESR continues to push nuclear mass measurements towards the neutron drip line.  
For example, a recent measurement of the $^{129-131}$Cd nuclei near the N=82 waiting point was
made with a mass accuracy $\delta m/m = 2\times 10^{-4}$ \cite{knobel16}.  From the perspective
of the astrophysical r process, this is useful for determining the strength of the N=82
shell gap.  Also, operating in IMS mode, mass measurements of the neutron-rich nuclei between
Se and Ce were made using a fragmented $^{238}$U primary beam \cite{knobel16b}.  This measurement
was important in extending the landscape of known nuclear masses by one additional neutron
past stability.  Recent 
measurements were also made for neutron-rich nuclei near Lu and Os in SMS 
mode \cite{attallah02}. Other measurements include 35 mass measurements of 
neutron-rich nuclei between As and Xe \cite{sun08}.
While many of the measured masses are not yet on the r-process path, they
are useful in constraining mass models near the rare earth region.  This is important as many 
mass models have significant disagreement in this region.
\subsubsection{HIRFL-CSRe} 
A newer facility located at the Heavy Ion Research Facility in Lanzhou (HIRFL)\cite{zhan10,yuan13,chen15,sun18,xing15,tu18}, the cooler-storage-ring (HIRFL-CSRe) has begun operation
in the last decade with mass measurements of over 60 nuclei and 20 first time 
measurements \cite{xu15,zhang17,zhang12,yan13,xu16,xing18,tu11,shuai14}.  Both proton-rich
and neutron-rich nuclei were re-measured or measured for the first time.  Of these, a significant number
of proton-rich $pf$-shell nuclei were measured for the first time at the CSRe with 
mass uncertainties $\delta m/m\sim$10$^{-7}$ -- 10$^{-6}$.  These measurements were
made by running the CSRe in IMS mode.  These are of particular interest for
evaluating the effects of nuclear structure and electron-capture properties of nuclei relevant to
type Ia supernovae.  In addition, as new masses are measured on or near the proton drip line, 
the structure of nuclei relevant to the astrophysical rp process can be better understood.

\begin{figure}
	\includegraphics[width=\textwidth]{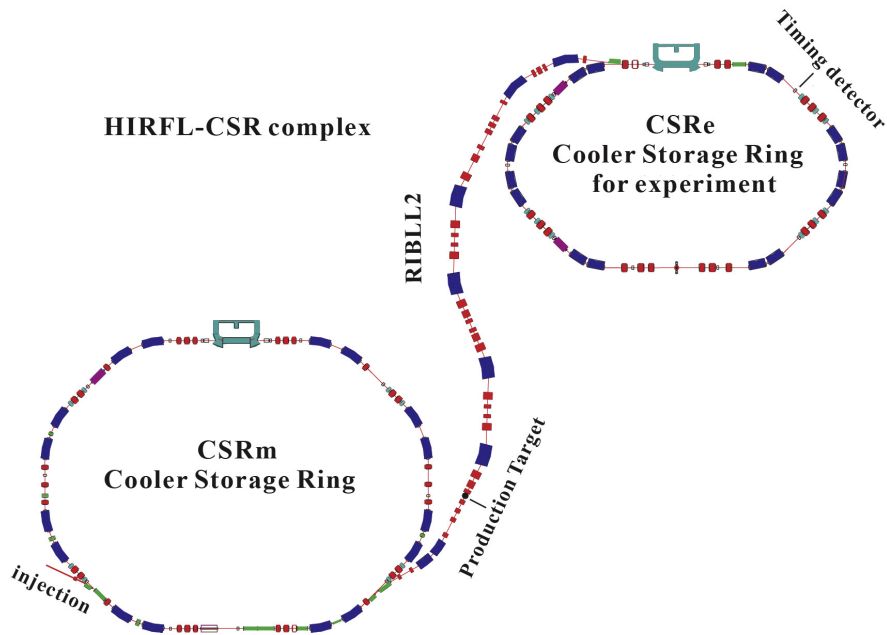}
	\caption{\label{csre_beam}Layout of the HIRFL-CSRe system, including the spectrometer to select ions, the production target, and the storage ring \cite{xu16}. The electron cooler is shown at the top of of the CSRe in the figure.  Dual timing detectors can be used in the straight section of the ring opposite the cooler.  Used with permission \cite{xu16}.}
\end{figure}

The CSRe is similar in structure to the GSI-ESR.  An example of the complete
beamline, including the injection system, is shown in Figure \ref{csre_beam} \cite{xu16}.  Secondary beams are produced from a primary beam with energies of
350 -- 480 MeV/u incident on a 10 -- 15 mm Be production target \cite{sun18}.  Isotopic species are
then selected in the RIBLL2 fragment separator \cite{sun03} as shown.  In addition to the timing detector shown in the
figure, the CSRe is able to employ two additional TOF detectors consisting of 19 $\mu$g/cm$^2$ carbon foils
which generate secondary electrons focused onto a micro-channel plate \cite{mei10}.  These two detectors
have been installed in the straight section of the ring opposite the electron cooler (shown in the figure). The nominal circumference of the CSRe is 129 m, and the magnetic rigidity can reach 8.4 Tm \cite{xu16}.  The use of the dual detectors is to correct for the velocity spread of the particles in the ring \cite{sun18}.

The HIRFL facility has also done a significant amount of work in correcting for magnetic field drift 
over the course of an experiment, providing for a significant improvement in time and frequency 
measurements \cite{xu11}.
\subsubsection{RIKEN Rare RI-Ring}
The Rare RI-Ring (R3) is located in Saitama, Japan. It is a 60 m circumference ring with a 1\% 
momentum acceptance \cite{ozawa12}.  Heavy beams with energies $\sim$ 200 MeV/u are injected, resulting
in a revolution time of about 355 ns.

Nuclei are separated by the BigRIPS facility from the primary target between a ring cyclotron and the
BigRIPS.  This facility was commissioned in 2015.  Because the RIKEN Radioactive Ion Beam Factory (RIBF)
can produce a very high intensity $^{238}$U beam, more exotic nuclear species can be produced at higher
intensities.  One interesting aspect of the R3 device is that particle ID is performed in the BigRIPS facility, and no identification is necessary within the ring itself.
Plans at R3 are to measure nuclear masses in IMS mode with a resolution of $\delta m/m\sim 10^{-6}$.
\subsection{Other Techniques}
\label{other_methods}
In addition to traps, rings, and TOF methods, other systems very similar to those previously mentioned
are worth discussing.  In particular, multi-reflection TOF (MR-TOF) mass measurement techniques as well
as some innovative methods exist.  A few additional techniques are described here.

\subsubsection{MR-TOF}
With the Multi-reflection time-of-flight method, ions are reflected back and forth between two
electrostatic mirrors in a short tube \cite{dickel13}.  This effectively takes a long flight
path and ``folds'' it several times into a short device.  The ions are then ejected 
from the device to be incident onto a timing detector.  Several measurements can be made
in which the number of reflections is varied, which can provide an indication of the time
for one passage of the particle in the device.

As an example, consider the KEK-MRTOF system, located in Japan, has proven to be quite useful
for nuclear mass measurements.  It is capable of measuring masses
with a resolution $\delta m/m\sim 10^{-7}$ at 
high rates and a measurement cycle of $\approx$30 ms \cite{schury17,schury18}.

\begin{figure}
	\includegraphics[width=\textwidth]{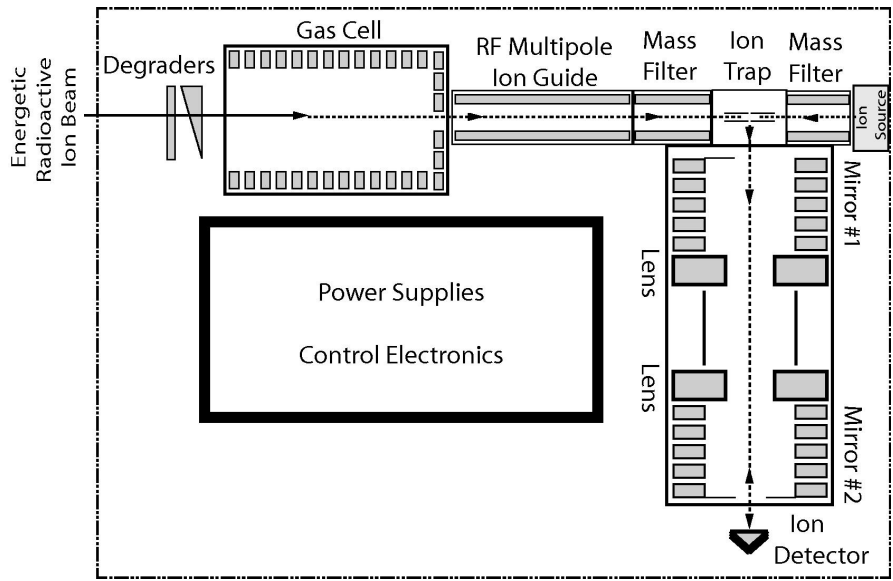}
	\caption{\label{kek_fig}Schematic diagram of the KEK-MRTOF \cite{schury13}.  After ions
		enter the ion trap, they are injected into the spectrometer (right side of 
		the figure).  Used with permission \cite{schury13}.}
\end{figure}

A schematic diagram of the setup of the KEK-MRTOF is shown in Figure \ref{kek_fig}\cite{schury13}.
An MRTOF spectrometer first traps ions.  After this, it injects those ions into a drift chamber in which
they are reflected back and forth via electrostatic mirrors.  The time spent in the drift tube
prior to ejection depends on the oscillation RF frequency of the electrostatic mirrors and the ion mass.
The ions are then ejected from the tube where a timing detector is used to provide a measurement of flight
time in the tube.  The measured velocity resolution can be improved by tuning the RF frequency of the
electrostatic mirror to increase the ion flight time within the limitations of its lifetime.

The KEK-MRTOF has been used to provide first-time mass measurements of 
rare-earth nuclei with 204$\le$A$\le$206 \cite{schury17}.  

Other MRTOF systems have been installed at other facilities, including - but not limited to - CERN-ISOLDE \cite{wolf16}, GANIL \cite{chauveau16}, Notre Dame \cite{brodeur17},
Argonne \cite{hirsh16}, RIKEN \cite{schury14}
TRIUMF \cite{jesch15}, and
GSI \cite{plass13}.
\subsubsection{MISTRAL} 
\begin{figure}
	\includegraphics[width=\textwidth]{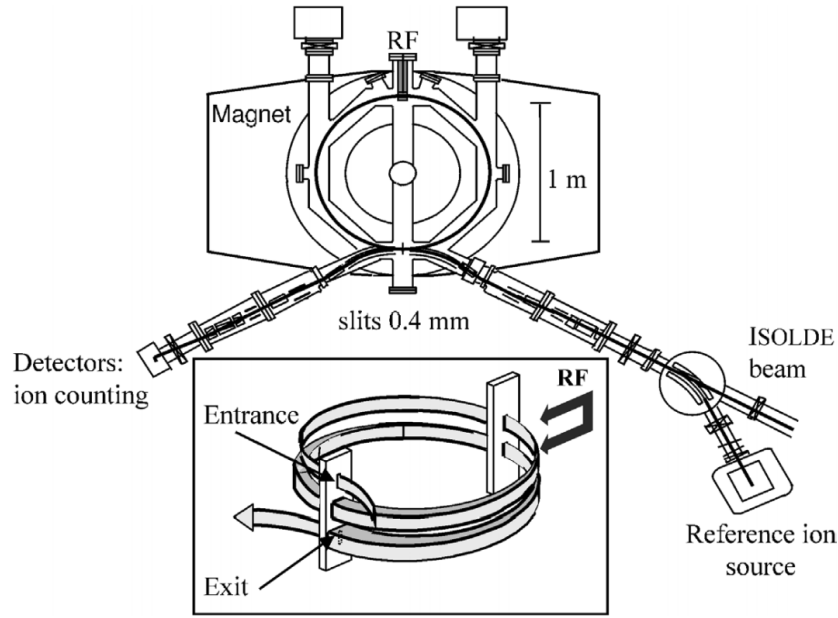}
	\caption{\label{mistral}Top: Drawing of the MISTRAL spectrometer.  Bottom: Schematic of 
		a particle path in MISTRAL \cite{gaulard06}.  Used with permission \cite{gaulard06}.}
\end{figure}

The MISTRAL device at ISOLDE is a frequency-based measurement system which is used
to measure the masses of nuclei via their cyclotron frequency in a uniform magnetic field \cite{lunney01}.
With this device, ions enter an RF modulator, in which their kinetic energy is determined.  After this, they 
enter a uniform magnetic field, where they travel with a cyclotron motion an simultaneous longitudinal motion
prior to extraction \cite{simon95}.  In order to be successfully extracted, they must undergo a specific number of 
cyclotron orbits.  Otherwise, they are not extracted.  Thus, the external field is directly related to ion mass.  A drawing of the MISTRAL device and a schematic of the path of a particle in MISTRAL is shown in 
Figure \ref{mistral} \cite{gaulard06}.

To date, MISTRAL has been used to measure the masses of light, neutron-rich nuclei, including $^{26}$Ne, $^{26 - 30}$Na, $^{29-33}$Mg, $^{11}$Li 
and $^{11,12}$Be  \cite{lunney01,gaulard06,bachelet08,gaulard09} with a resolution of $\sim$20 keV.  
\section{Mass Measurement Compilations and Evaluations}
Many existing mass compilations are now supplemented by online resources in
various formats which allow for easy access, searches, and analysis.  
As the field of nuclear physics continues to grow rapidly, with the volume of
data growing, these online resources can be extremely useful in searching for updated data
as well as for planning future experiments.

Among these
online resources are \texttt{nuclearmasses.org} \cite{nucmass} and the 
National Nuclear Data Center (NNDC) \cite{nndc}.  The \texttt{nuclearmasses.org} online
resources
contains a useful interface for comparing multiple datasets as well 
as up-to-date compilation of new measurements.  The NNDC provides
an interactive tool for examining the isotopic chart color-coded by multiple
properties.  In addition, it contains links to access other data sets.

The most recent Atomic Mass Evaluation, AME2016, \cite{ame16a,ame16b} replaces the former
AME03 \cite{ame03,ame03a,ame03b}.  The online volume also provides user-friendly
access to the NUBASE2016 tables \cite{nubase16,ame16a,ame16b} in various formats.  
This compilation also provides evaluations and evaluation procedures of extrapolated masses
and properties.  The NUBASE2016 tables can be accessed in electronic form from multiple
online resources \cite{amdc_ch,amdc_au}.

In addition to recent mass evaluations, the Atomic Mass Compilation 2012 (AMC12) \cite{amc12} 
contains evaluated and experimentally determined masses, which is useful in that the 
tabulated results indicate the facility from which the results come.  Comparisons
between AMC12 and AME03 have been made in additional resources \cite{audi15}.
\section{Future Work}
There is still a significant portion of the table of isotopes yet to be explored.  
Probing the drip lines and verifying particle stability of nuclei is perhaps
the most fundamental nuclear property to probe.  Beyond this, perhaps the
second-most fundamental property is nuclear masses, though this is not necessarily
the easiest to measure.  The field of measuring nuclear masses is far from
closed, and multiple directions and facilities are being developed.

New opportunities and ideas have arisen in the past few years.  Notably, studies
of superheavy elements, the discovery of new elements, and new developments
in the field of multi-messenger astronomy have made significant progress. These fields, among others,
have provided additional impetus to continue to probe the table of isotopes.

From an astrophysical standpoint, nuclear masses are fundamentally important in
understanding specific astrophysical processes.  For example, as mentioned
in the introduction, masses of nuclei along the r-process path are useful
in extracting neutron separation energies and shell structure, thus informing
further theoretical development.

Multiple facilities are currently being developed or upgraded to address - among other things - the 
needs of nuclear mass measurement programs.  
Perhaps the greatest need for new facilities is the need to produce more exotic nuclei with 
production rates that are accessible with mass measurement programs and techniques.
Some of the newest facilities under development are listed here.  
These are all in various stages of development.  
\subsection{FRIB}
The Facility for Rare Isotope Beams (FRIB) is nearing completion with an anticipated commissioning
date of around 2022.  This facility represents a major reconstruction of the NSCL, described in 
\S\ref{nscl_desc} with the predicted rare isotope production rates shown in Figure \ref{frib_rates}.

While the scientific scope of the FRIB facility is broad, its use measuring nuclear masses
has not been overlooked \cite{surman18,gade16}.  With the expected capabilities of this facility, 
nuclear mass measurements using the existing LEBIT facility \cite{ringle09} may be possible.  For nuclei
beyond the reach of LEBIT, measurements via the B$\rho$-TOF method are possible.  To address this
method, the High Rigidity Spectrometer (HRS) is being developed \cite{baumann16}.
\subsection{GSI-FAIR}
The Facility for Antiproton and Ion Research at GSI (GSI-FAIR) began construction in summer 2017.  It will
use existing facilities at GSI as input stages to the designed accelerator complex \cite{kester}.  This 
facility is expected to provide high-intensity beams covering a range of species, including uranium \cite{barth17}. 

Nuclear mass measurements at the upgraded facility are planned as part of the NuSTAR project \cite{munzenberg13,gerl}.  Masses measured at GSI-FAIR will benefit from beam intensities of
one to a few orders of magnitude higher than existing GSI intensities.  Nuclear masses
can be measured with improved techniques including the use of storage rings at GSI \cite{NESR}.
\subsection{RAON}
In Korea, the Rare isotope Accelerator complex for ON-line experiment (RAON) began construction in
2018 with an estimated completion date in 2021 \cite{jeong18}.  This facility is planned to have both  isotope separation online (ISOL) and an in-flight fragmentation capabilities.  

The vision for the planned facility includes a MR-TOF system to achieve mass resolutions of 
$\sim$10$^{-6}$.  The separator is predicted to produce some very neutron-rich nuclei.
Masses of very exotic nuclei, possibly as neutron-rich as $^{60}$Ca may be possible
when the facility comes online \cite{jeong18}.   
\subsection{HIAF}
The High Intensity heavy-ion Accelerator Facility (HIAF) is the planned next-generation facility to
upgrade the existing HIRFL facility in China \cite{yang13,wu18}. The first beam from HIAF is planned in 2024.

For nuclear mass measurements, the Spectrometer Ring (SRing) is planned \cite{wu18}, a 15 Tm ring with a 
circumference of 273.5 m and excellent momentum resolution.  
\section{Conclusions}
As new online and planned radioactive ion beam facilities push production rates of exotic nuclei to 
higher values, the field of nuclear mass measurements continues to grow.  Production rates of even more
exotic
nuclei are becoming high enough to make these nuclei accessible by various techniques.   While the lifetimes of 
some of the very neutron-rich nuclei are so short that some techniques are excluded, it may still be
possible to measure nuclear masses with existing TOF methods (see Figure \ref{mass_params}).

TOF methods are currently being accompanied by development of high-rigidity spectrometers, some of which
are already in use, such as the SHARAQ spectrometer \cite{michimasa13,michimasa16}.  Others are currently
being planned, such as the FRIB HRS \cite{baumann16}.  These are outfitted with beamline detectors for
high-precision determination of particle rigidity, which is necessary for precise event-by-event measurements
as well as separation of charge states.

The new facilities and improvements in existing techniques will result in a better understanding of mass
models far from stability, evolution of shell structure, and better understanding of astrophysical processes
in the next half decade.

Future mass measurements will be useful in informing theoretical developments as
well.  As computational power improves, the ability for larger basis sets
in shell model calculations can result in more precise predictions.  As
more advanced mass models become available, these can be tested and better
constrained with additional measurements.  While much of mass measurement
experimentation centers on masses of astrophysical interest and more exotic
masses, it is still useful to push the limits of known nuclear masses even one
or two mass units at a time because many existing mass models disagree 
significantly outside the regions of known nuclear masses.

\newcommand{\actaa}{Acta Astron. }%
\newcommand{\araa}{Ann.Rev.Astron.\&Astroph. }%
\newcommand{\apj}{Astroph.J. }%
\newcommand{\apjl}{Astroph.J.Lett. }%
\newcommand{\apjs}{Astroph.J.Supp. }%
\newcommand{\ao}{Appl.~Opt. }%
\newcommand{\apss}{Astroph.J.\&Sp.Sci. }%
\newcommand{\aap}{Astron.\&Astroph. }%
\newcommand{\aapr}{Astron.\&Astroph.~Rev. }%
\newcommand{\aaps}{Astron.\&Astroph.~Suppl. }%
\newcommand{\aj}{Astron.Journ. }%
\newcommand{\azh}{AZh }%
\newcommand{\gca}{GeoCh.Act}%
\newcommand{\memras}{MmRAS }%
\newcommand{\mnras}{Mon.Not.Royal~Astr.~Soc. }%
\newcommand{\na}{New Astron. }%
\newcommand{\nar}{New Astron. Rev. }%
\newcommand{\pra}{Phys.~Rev.~A }%
\newcommand{\prb}{Phys.~Rev.~B }%
\newcommand{\prc}{Phys.~Rev.~C }%
\newcommand{\prd}{Phys.~Rev.~D }%
\newcommand{\pre}{Phys.~Rev.~E }%
\newcommand{\prl}{Phys.~Rev.~Lett. }%
\newcommand{\pasa}{PASA }%
\newcommand{\pasp}{Proc.Astr.Soc.Pac. }%
\newcommand{\pasj}{Proc.Astr.Soc.Jap. }%
\newcommand{\rpp}{Rep.Prog.Phys. }%
\newcommand{\skytel}{Sky\&Tel. }%
\newcommand{\solphys}{Sol.~Phys. }%
\newcommand{\sovast}{Soviet~Ast. }%
\newcommand{\ssr}{Space~Sci.~Rev. }%
\newcommand{\nat}{Nature }%
\newcommand{\iaucirc}{IAU~Circ. }%
\newcommand{\aplett}{Astrophys.~Lett. }%
\newcommand{\apspr}{Astrophys.~Space~Phys.~Res. }%
\newcommand{\nphysa}{Nucl.~Phys.~A }%
\newcommand{\physrep}{Phys.~Rep. }%
\newcommand{\procspie}{Proc.~SPIE }%

\bibliographystyle{ws-ijmpe}

\end{document}